\documentclass{article}
\usepackage{spconf,amsmath,graphicx}
\usepackage{booktabs,caption}
\usepackage[normalem]{ulem}
\usepackage[flushleft]{threeparttable}
\useunder{\uline}{\ul}{}
\usepackage{subfig}

\title{Revisiting SVD to generate powerful Node Embeddings for Recommendation Systems}

\name{Amar Budhiraja}
\address{Myntra Designs Pvt. Ltd.}

\begin{document}
%
\maketitle
\begin{abstract}
Learning node representations is at the heart of Graph Representation Learning (GRL), and has been proven successful in establishing the state-of-the-art in multiple domains including recommendation systems. In this paper, we benchmark node embeddings generated through Singular Value Decomposition (SVD) of adjacency matrix for embedding generation of users and items and use a two-layer neural network on top of these embeddings to learn relevance between user-item pairs. Inspired by the success of higher-order learning in GRL, we further propose an extension of this method to include two-hop neighbors for SVD through the second order of the adjacency matrix and demonstrate improved performance compared with the simple SVD method which only uses one-hop neighbors. Empirical validation on three publicly available datasets of recommendation system demonstrates that the proposed methods, despite being simple, beat many state-of-the-art methods and for two of three datasets beats all of them up to a margin of 10\%. Through our research, we want to shed light on the effectiveness of matrix factorization approaches, specifically SVD, in the deep learning era and show that these methods still contribute as important baselines in recommendation systems.
\end{abstract}

\section{Introduction}
Graph Representation Learning (GRL) presents a very promising direction in terms of machine learning on graphs \cite{grover2016node2vec,kipf2016semi,van2017graph,wang2019heterogeneous}. A central idea in GRL is to represent each node in the graph as a vector of floating-point numbers to capture some desired properties of a node with respect to the graph. For node2vec\cite{grover2016node2vec}, this property is the neighborhood of the node; for GraphSage\cite{hamilton2017inductive}, it is about capturing the feature neighborhood of a node based on a selected neighbor set; for Graph Convolutional Network (GCN) \cite{kipf2016semi}, the purpose of a node embedding is to capture both feature and neighborhood similarity. These methods have been profoundly useful in several domains such as bio-inspired machine learning and genomics \cite{gonzalez2021predicting, laponogov2021network, wu2021bridgedpi}, spam detection \cite{liu2018heterogeneous, qu2020category}, natural language processing \cite{ragesh2021hetegcn,yao2019graph,zhang2020every,linmei2019heterogeneous} and recommendation systems \cite{wang2019neural, zhao2017meta,he2020lightgcn, sunmulti, zhao2019t}.

In recommendation systems, GRL has been applied to further advance collaborative filtering algorithms by considering multi-hop relationships between users and items \cite{wang2019neural}. The authors in \cite{wang2019neural} further proposed the notions of message dropout and node dropout to reduce overfitting in GCN like methods. In a follow-up study \cite{he2020lightgcn}, it was demonstrated that simplifying GCN network by reducing non-linearity from the network can give a boost to the performance of these higher-order methods. Their work also corresponded with a similar study done in \cite{wu2019simplifying} where the authors argued that for GCN, even after removing non-linearity and collapsing weight matrices into a single one, the performance does not degrade in downstream tasks. The research carried out in the above papers compares several state-of-the-art methods to the proposed methods and shows that the simplicity of models is leading to higher performance, credited to better generalization of the models. Motivated by these studies, we set out to benchmark a simple SVD based approach in this paper on the recommendation systems problem to understand if further simplicity of the modelling approach can improve the performance metrics. In the proposed method, we first generate user and item embeddings using SVD of the adjacency matrix of the user-item interaction graph and then employ a two-layer neural network with these embeddings as inputs to estimate the relevance of an item to a user.

Given the success of multi-hop graph neural network models in previous studies, we augment the simple SVD method to consider a two-hop adjacency matrix for generating the embeddings and found that this method outperforms the simpler one-hop SVD method as well. Empirical results on three public datasets demonstrate that the performance of the proposed methods is indeed comparable to state-of-art approaches, and these methods beat many of them despite their simplicity. For two out of three datasets, the methods even outperform all compared approaches and effectively establish new state-of-the-art performance with the margin of improvement as much as 10\%. 

The rest of the paper is divided into three sections: Section 2 describes the proposed methods, Section 3 contains the empirical experiments, and Section 4 provides the conclusion and future work.

\section{Proposed Methods}
In this section, we elaborate on the proposed methods. We first discuss the SVD based baseline followed by an extension of the same by using two-hop matrices. In the last part, we describe the loss function and model training. Before discussing the proposed methods, we list our notations in Table \ref{notation}.

\begin{table*}[]
\centering
\begin{tabular}{l|l}
\toprule
\textbf{Symbol} & \textbf{Definition}    \\ \hline
$\mathcal{A}$ & Adjacency matrix between Users and Items     \\ \hline
$\mathcal{A}'$    & Symmetric adjacency matrix between Users and Items   \\ \hline
$\widetilde{\mathcal{A}}$    & Laplacian Normalization of matrix $\mathcal{A}'$   \\ \hline
D    & Degree Matrix of $\mathcal{A}'$  \\ \hline
\textit{u}    & User  \\ \hline
\textit{i}  & Item     \\ \hline
$e_{u}^{SVD}$   & Embedding of user generated from SVD of $\widetilde{\mathcal{A}}$   \\ \hline
$e_{i}^{SVD}$       & Embedding of item generated from SVD of $\widetilde{\mathcal{A}}$   \\ \hline
$e_{u}^{mj}$           & Embedding of user outputted from the $j^{th}$ layer of perceptron model \\ \hline
$e_{i}^{mj}$        & Embedding of item outputted from the $j^{th}$ layer of perceptron model \\ \hline
$e_{u}$     & Concatenation of : $e_{u}^{SVD}$ and $e_{u}^{mj}$ (j = $\{1,2\}$)         \\ \hline
$e_{i}$     & Concatenation of : $e_{i}^{SVD}$ and $e_{i}^{mj}$  ( j = $\{1,2\}$)       \\ \bottomrule
\end{tabular}
\caption{Symbols and their meaning used in the paper. \label{notation}}
\end{table*}

\subsection{Simple SVD Baseline}

Matrix factorization is a well-studied problem in linear algebra and has been extensively applied to recommendation systems \cite{ungar1998clustering, li2019improved, debnath2008feature, choi2012hybrid}, typically in the form of collaborative filtering. In this paper, we propose a simple approach to generate user and item embeddings using Single Value Decomposition (SVD) \cite{klema1980singular} of the adjacency matrix between users and items. Using a two-layer perceptron model, we transform these embeddings in a supervised fashion to learn the relevance between the user and item pairs. We call this method \textbf{S}imple \textbf{S}VD \textbf{B}aseline (SSB).

To compute the SVD embeddings, we consider the adjacency matrix of the user-item interaction graph, $\mathcal{A}$. We first convert the asymmetric matrix to a symmetric matrix as follows:

$$
\mathcal{A}' = 
\begin{pmatrix}
  \begin{matrix}
  0
  \end{matrix}
  &  \mathcal{A} \\
  \mathcal{A}^T  &
  \begin{matrix}
  0
  \end{matrix}
\end{pmatrix}.
$$

\begin{figure}
 \includegraphics[scale=0.4]{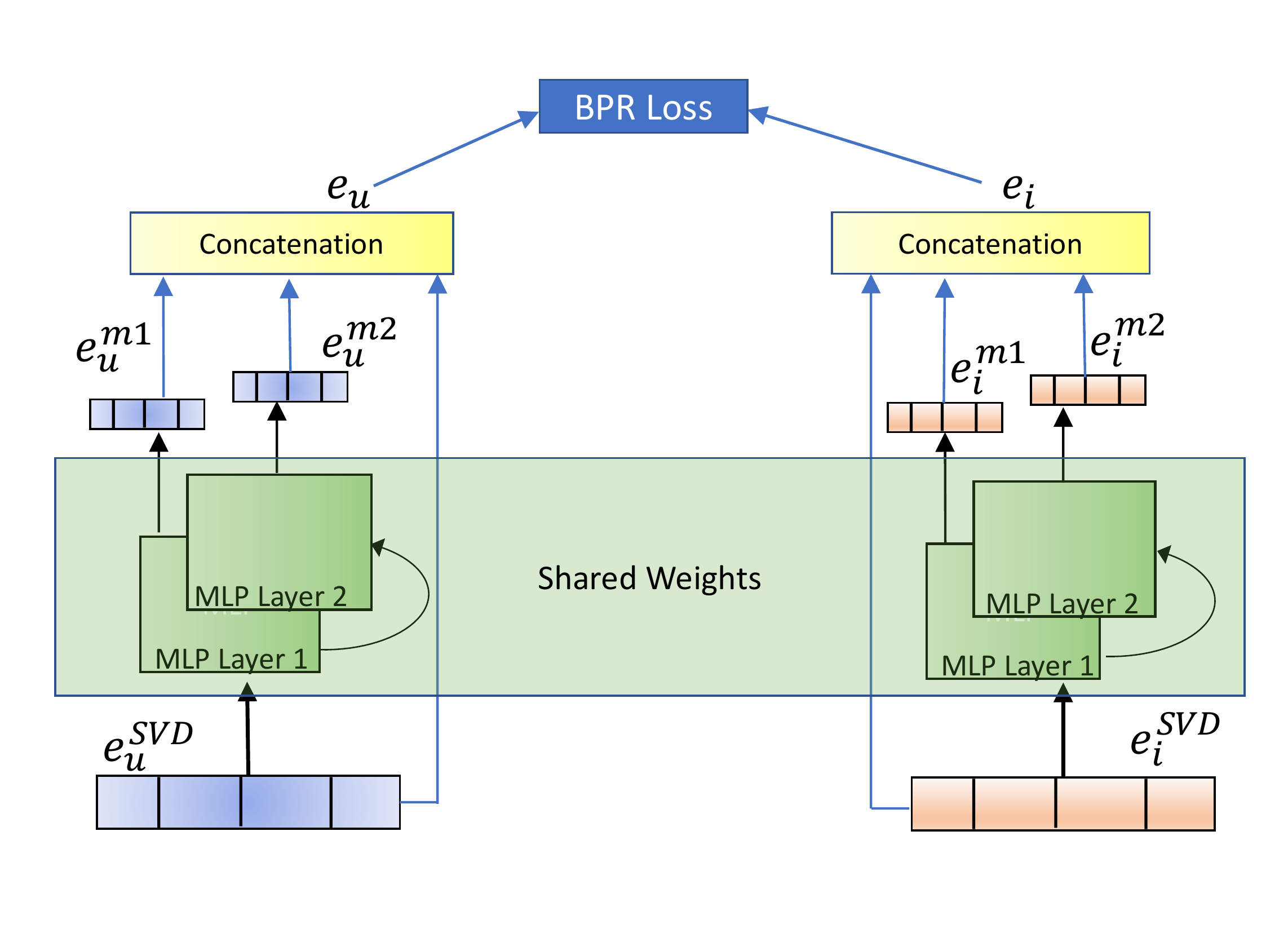} \caption{Model architecture for SSB. $e_{u}^{SVD}$ and $e_{i}^{SVD}$ are the user embedding and item embedding respectively generated from the Truncated SVD \cite{hansen1990truncated} of $\widetilde{\mathcal{A}}$. $e_{u}^{m1}$ and $e_{u}^{m2}$ are the embedding outputs from the first and second layers of MLP respectively. Finally, $e_{u}^{SVD}$, $e_{u}^{m1}$ and $e_{u}^{m2}$ are concatenated together to form the user embedding $e_{u}$. The item embedding $e_{i}$ is constructed in the same way as $e_{u}$. Dot product between $e_{u}$ and $e_{i}$ is used the score for the user-item pair and the same is optimized through backpropagation using pairwise BPR loss similar to the previous studies\cite{wang2019neural, he2020lightgcn}. \label{arch}}
\end{figure}

We then compute a Laplacian Normalization of $\mathcal{A}'$ as discussed in \cite{kipf2016semi}: $ \widetilde{\mathcal{A}} = \mathcal{D}^{-\frac{1}{2}} \mathcal{A}'\mathcal{D}^{-\frac{1}{2}} $,
where $\widetilde{\mathcal{A}}$ is the Laplacian Normalized of adjacency matrix, $\mathcal{A}'$, and $D$ is the degree matrix derived from $\mathcal{A}'$.

We perform matrix factorization on $\widetilde{\mathcal{A}}$ using Truncated SVD on this normalized matrix to generate user embeddings ($e_{u}^{SVD}$) and item embeddings ($e_{i}^{SVD}$), where the number of components in Truncated SVD correspond to the embedding dimension. We use Truncated SVD \cite{hansen1990truncated} since it has shown to be scalable on large matrices.

After generating these embeddings, we transform them through a two-layer perceptron model ($f(x) = x$, as the activation function) and concatenate the output of both the layers of the perceptron model along with original SVD embedding to generate a user embedding ($e_{u}$) or an item embedding ($e_{i}$). The intuition behind using the perceptron model is to allow supervised transformation of $e_{u}^{SVD}$ and $e_{i}^{SVD}$ to learn the relevance between user and item. Fig. \ref{arch} shows the model architecture. The dot product between $e_{u}$ and $e_{i}$ acts as the relevance score for the user-item pair and is optimized by the model through tuning of the weights of the two-layer perceptron model via back-propagation.

\subsection{Two-Hop SVD Approach}
Motivated by the success of multi-hop graph neural networks and the performance of the SSB approach on recommendation tasks, we attempt at joining both of these into a single method to capture higher-order relationships between users and items, similar to graph neural networks like GCN \cite{kipf2016semi}. 

The overall model architecture remains the same as SSB, except for the change in how the $e_{u}^{SVD}$ and $e_{i}^{SVD}$ embeddings are computed. To compute an embedding that can capture the two-hop signals, we compute the second power of the Laplacian Normalized adjacency matrix, $\widetilde{\mathcal{A}}$, and then compute its Truncated SVD. We finally concatenate the embeddings from SVD of $\widetilde{\mathcal{A}}$ (corresponding to one-hop neighborhood) and SVD of $\widetilde{\mathcal{A}}^{2}$ (corresponding to two-hop neighborhood) to generate $e_{u}^{SVD}$ and $e_{i}^{SVD}$ embeddings for this approach. The embedding size of TSA is the size of the vector after this concatenation. Since this approach contains two-hop signals from the graph, we denote this method as \textbf{T}wo-Hop \textbf{S}VD \textbf{A}pproach (TSA).

\subsection{Model Training}

The learnable parameters in the proposed methods are only the weights of the multi-layer perceptron model. To optimize the user-item relevance, we employ the Bayesian Personalized Ranking (BPR) loss \cite{rendle2012bpr} similar to \cite{wang2019neural}. It is a pair-wise loss that encourages correct predictions on observed instances than on unobserved instances. We use the Adam optimizer \cite{kingma2014adam} in a mini-batch setting, where the batch size is a hyperparameter.

\section{Experiments}


\subsection{Datasets and Performance Metrics}

We use the same three datasets (Gowalla, Yelp2018 and Amazon-Book) as \cite{wang2019neural, he2020lightgcn} with the same train and test split in order to make a fair comparison with the already reported results. Table \ref{stats} summarizes the dataset statistics. We refer the reader to \cite{wang2019neural} for more details of the datasets. We evaluate the performance on mean NDCG@K and mean Recall@K per user for K=20. We keep K=20 to enable fair comparison with previous studies which use the same metrics \cite{wang2019neural, he2020lightgcn}. For the rest of the paper, we denote Recall@20 as Recall and NDCG@20 as NDCG. It should be noted that for both Recall and NDCG, the items retrieved for top-20 are solely from the test partition of the dataset.

\subsection{Hyperparameter Tuning}
For the proposed methods, there are four key hyperparameters - SVD embedding size ($|e_{u}^{SVD}|$ and $|e_{i}^{SVD}|$), batch size,  learning rate and size of the multi-layer perceptron. For this study, we fix the size of the multi-layer perceptron to be 512 neurons each and keep the learning rate as $10^{-3}$ motivated by the experiments in \cite{he2020lightgcn}. We tune the SVD embedding size over the following set: $\{ 2^6, 2^7, 2^8, 2^9, 2^{10}\}$. We keep the batch size as 1024 for Gowalla and Yelp2018, and 2048 for Amazon-Book as used in the study of NGCF \cite{wang2019neural} and LightGCN \cite{he2020lightgcn}.

\begin{table}
\centering
\begin{tabular}{c|c|c|c|c}
\toprule  \textbf{Dataset}      & \textbf{\#Users}     & \textbf{\#Items}  & \textbf{\#Interactions} & \textbf{Density}  \\ \midrule
Gowalla & 29,858 & 40,981 & 1,027,370 & 0.00084 \\
Yelp2018 & 31,688 & 38,048 & 1,561,406 & 0.00130 \\
Amazon-Book  & 52,643 & 91,599 & 2,984,108 & 0.00062 \\ \bottomrule
\end{tabular}
\caption{Dataset Statistics\label{stats}}
\end{table}

\subsection{Empirical Results}

\subsubsection{\textbf{Comparison with state-of-the-art methods}}
$\newline$
In this section, we report the performance metrics for the proposed methods - Simple SVD Baseline (SSB) and Two-Hop SVD Approach (TSA). We benchmark the approach against NGCF \cite{wang2019neural}, Mult-VAE\cite{liang2018variational}, GRMF \cite{rao2015collaborative}, LightGCN \cite{he2020lightgcn}, MF \cite{rendle2012bpr}, and NeuMF \cite{he2017neural}. Although MF \cite{rendle2012bpr}, and NeuMF \cite{he2017neural} methods are relatively older methods to compare. However, we report their performance here as these are closely related to matrix factorization in the context of recommendation systems. LightGCN \cite{he2020lightgcn} is the state-of-the-art method showing the best performance compared to all related approaches as shown in their paper. Table \ref{results} shows the performance metrics for the proposed methods and the compared methods. We replicate the results of these approaches from the original papers of NGCF \cite{wang2019neural} and LightGCN \cite{he2020lightgcn}. We follow the same experimental methodology as stated in the papers and followed in the code and datasets made available by the authors of these studies to make a fair comparison.

\begin{table*}[]
\centering
\begin{tabular}{@{}lllllllll@{}}
\toprule
Dataset          & \multicolumn{2}{l}{Gowalla}      &  & \multicolumn{2}{l}{Yelp2018}      &  & \multicolumn{2}{l}{Amazon-Book} \\ \midrule
    & Recall         & NDCG            &  & Recall          & NDCG            &  & Recall         & NDCG           \\ \midrule
MF$^{\phi}$        & 0.1291         & 0.1109          &  & 0.0433          & 0.0354          &  & 0.0250         & 0.0196         \\
NeuMF$^{\phi}$       & 0.1399         & 0.1212          &  & 0.0451          & 0.0363          &  & 0.0258         & 0.0200         \\ \midrule
NGCF$^{\psi}$     & 0.157          & 0.1327          &  & 0.0579          & 0.0477          &  & 0.0344         & 0.0263         \\
Mult-VAE$^{\psi}$   & 0.1641         & 0.1335          &  & 0.0584          & 0.0450          &  & 0.0407         & 0.0315         \\
GRMF$^{\psi}$       & 0.1477         & 0.1205          &  & 0.0571          & 0.0462          &  & 0.0354         & 0.0270          \\
LightGCN$^{\psi}$   & \textbf{0.183} & \textbf{0.1554} &  & 0.0649          & 0.0530          &  & 0.0411         & 0.0315         \\ \midrule
{\ul SBB} \space\space ($|e^{SVD}| = 512$) & 0.169          & 0.1401          &  & 0.0647          & 0.0534          &  &     0.0408           &  0.0325              \\
{\ul TSA} \space\space ($|e^{SVD}| = 1024$) & 0.1704         & 0.1415          &  & \textbf{0.0657} &  \textbf{0.0542} & 
 &\textbf{ 0.0456} & \textbf{0.0364} \\ \bottomrule
\end{tabular}

\caption{Comparison of the proposed methods - SBB and TSA with related methods. It can be observed that the proposed methods, despite being very simple, beat all the compared methods for the Yelp2018 and Amazon-Book datasets. For the Gowalla dataset, the proposed methods prove to be a strong baseline and was able to beat all but one state-of-the-art methods on this dataset. $|e^{SVD}|$ is the size of Truncated SVD embedding which is same for both users and items. $\psi$ Results reused from \cite{he2020lightgcn}; $\phi$ Results reused from \cite{wang2019neural}. \label{results}}
\end{table*}

We can observe that TSA performs considerably better for Amazon-Book dataset than all the compared state-of-art methods, including LightGCN and NGCF. The relative gain of TSA over LightGCN, which performs the best among the compared methods, is approximately 9.86\% in Recall@20 and 13.4\% in NDCG@20. The performance of SSB is also higher than all considered approaches in terms of NDCG and only 0.7\% short of LightGCN and still better than all compared approaches.

We can see that for Yelp2018 dataset, TSA performed marginally better than LightGCN \cite{he2020lightgcn} in terms of both Recall (~1\% relatively) and NDCG (~2\% relatively). In contrast, SBB performs marginally lower than LightGCN \cite{he2020lightgcn} in terms of Recall but has a slightly higher NDCG. However, both SBB and TSA performs significantly better than all other compared approaches, including NGCF \cite{wang2019neural} which uses multi-hop relations among users to exploit higher-order signals for predicting relevance.

For the Gowalla dataset, the proposed methods perform poorly compared to LightGCN. There is ~1.3\% absolute difference in Recall and  ~1.5\% absolute difference in NDCG. However, despite being fairly plain, the proposed approaches outperform all other approaches, including some which are inherently more complex such as Mult-VAE and NGCF. 

We believe that the generalization power of the proposed methods has led to the improvement in performance in Yelp2018 and Amazon-Book datasets. However, in the case of Gowalla, the method seems to be underfitting because the signals of SBB and TSA are only limited to one-hop and two-hop neighbors respectively. As shown in \cite{he2020lightgcn}, as the number of layers of GCN is increased to four (equivalent to four-hops of neighborhood), the performance increases. We leave this aspect of experimentation to be addressed in future work.

\subsubsection{\textbf{Comparison of SBB and TSA}}
$\newline$
In this section, we point out the comparison between SBB and TSA approaches. In Table \ref{results}, we can observe that TSA always performs better than SBB, given the additional signal from two-hop neighbors. Comparing the proposed methods, SBB and TSA, we only see a small relative increase in performance in Gowalla and Yelp-2018 and a more significant uplift in the case of TSA for Amazon-Book dataset. This is in line with the studies of LightGCN \cite{he2020lightgcn}, and NGCF \cite{wang2019neural} where it was shown that as the number of layers is increased for the GCN model (which is equivalent to covering more hops in neighborhood), the performance increases in equivalent amount. We believe the improvement of metrics in TSA is the addition of signal from two-hops away which is not present in SBB. Interestingly, for Yelp2018 and Amazon-Book datasets, the two-hop proposed approach, TSA, is able to outperform the four-hop approaches of LightGCN and NGCF. 

\begin{table}[]
\centering
\begin{tabular}{@{}lll@{}}
\toprule     & SSB & TSA \\ \midrule
Training Loss   & 0.04477    &  0.03847    \\
Training Recall  & 0.14736    &  0.16262   \\
Training NDCG & 0.25875    &  0.28243   \\
\begin{tabular}[c]{@{}l@{}}Time for SVD \\ Computation\end{tabular} &  21.83s   &   503.04s  \\ \bottomrule
\end{tabular}
\caption{Comparison of the proposed methods with each with respect to training metrics on Yelp-2018 dataset. It should be noticed that SVD is only performed once at the start of training the model.\label{sbb_tsa_train}}
\end{table}

\begin{figure*}%
    \centering
    \subfloat[\centering Training Loss per epoch for Yelp2018 datatset]{\includegraphics[scale=0.095]{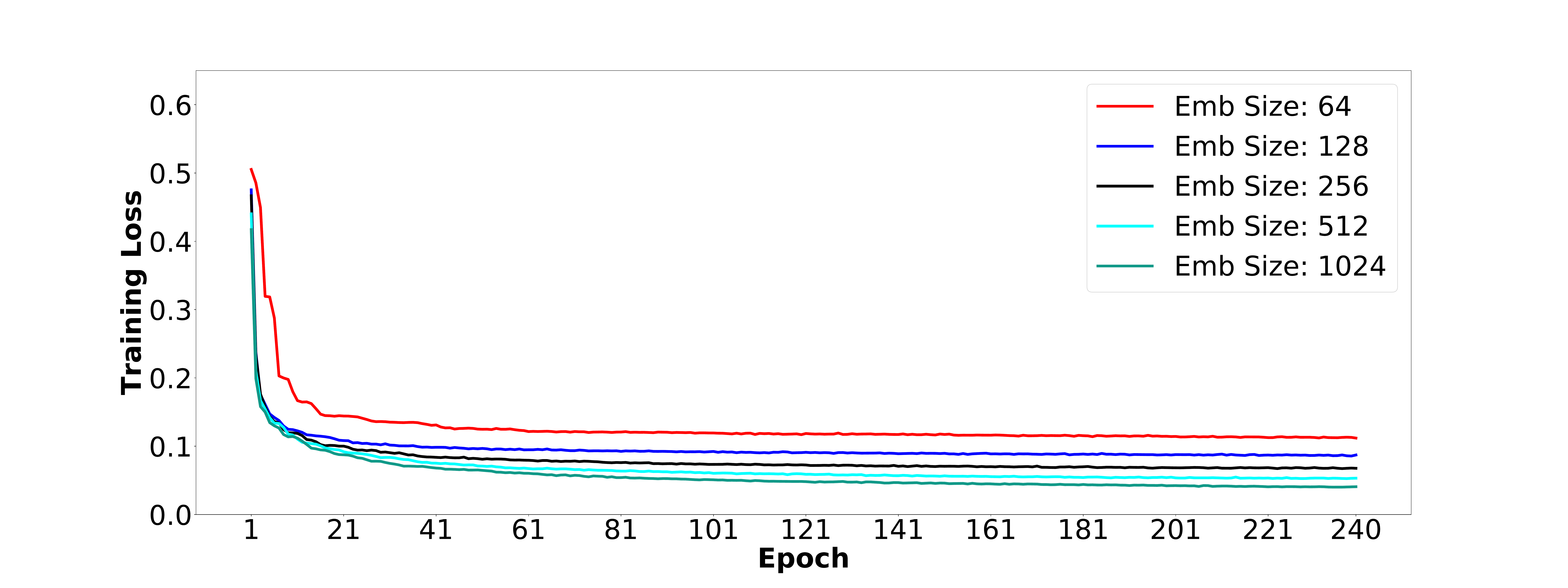} }%
    \subfloat[\centering Test Recall@20 per epoch for Yelp-2018 ]{
    \includegraphics[scale=0.09]{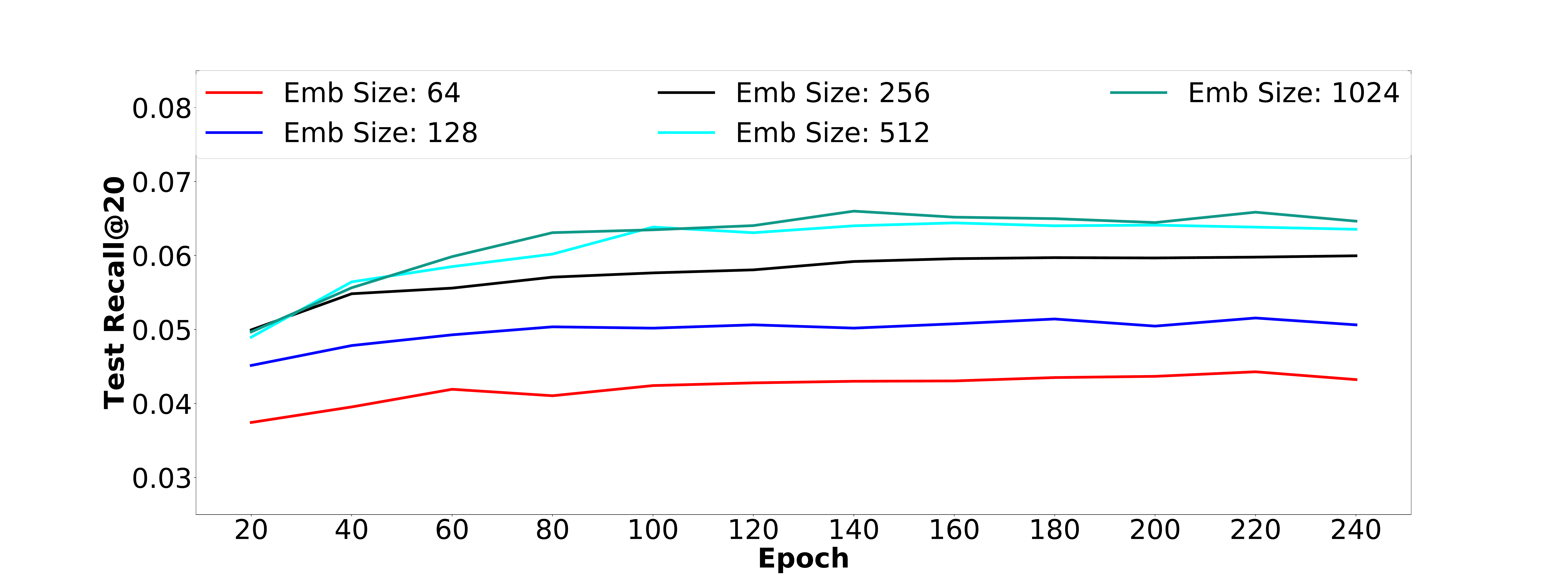} }%
    \caption{Training Loss and Test Recall@20 per epoch for the Yelp2018 for TSA method. For higher embedding dimension, Training Loss is lower and Test Recall is higher. }%
    \label{ablation}%
\end{figure*}

Regarding training loss, we observe that SSB, on convergence, has a higher training loss than TSA.  While TSA performs better than SSB, it does so at the expense of additional time to compute the two-hop adjacency matrix; and the additional Truncated SVD on the two-hop matrix. In Table \ref{sbb_tsa_train}, we summarize the comparison between SSB and TSA on the training metrics, while test performance is shown in Table \ref{results}. We run our experiments on \textit{Intel(R) Xeon(R) CPU E5-2690 v4 @ 2.60GHz} with 6 cores and 110 Giga Bytes of RAM.

\subsubsection{\textbf{Impact of Embedding Size on Performance on TSA}}
$\newline$
Out of the four hyperparameters discussed, we only optimize the SVD based embedding size for users and items, which become the input to the multi-layer perceptron. In this section, we will describe the observations on how the performance and training loss changes as we change the embedding dimensions for TSA, which performs better than SSB across datasets. We vary the embedding dimension as follows: $\{ 2^6, 2^7, 2^8, 2^9, 2^{10}\}$. Fig. \ref{ablation}(a) shows the training loss for different embedding sizes for the TSA approach.  As expected, it can be seen that as the embedding size increases, the loss decreases faster and also to a lower value. We observe a similar trend in test performance metrics, and Fig. \ref{ablation}(b) shows Recall@20 for the test set for different embedding sizes, and it can be seen that higher embedding sizes lead to better performance.


\section{Conclusions and Future Work}
In this paper, we started out to benchmark SVD based methods against the state-of-the-art GRL methods. We propose two approaches for the same, and experiments on three real-world open datasets demonstrate that these methods are powerful enough to beat many GRL methods and even come out as state-of-the-art themselves in two out three datasets. We observed the most significant relative gain of over 10\% against the state-of-the-art methods.

This particular work raises many research questions, and we envision the following future work. We plan on investigating how to generalize the approach from two-hop to n-hop since we saw in earlier research articles that with a higher order of neighborhood, we could expect better performance. There is also a need to propose an inductive version of these methods since transductive versions (as proposed in this paper), do not work with new nodes or new edges in the graph and would require frequent retraining in the current form. We also plan to investigate the aspects around the better implementation of SVD in big data frameworks such as Spark \cite{alexopoulos2020two,sun2015spark}. This would help us further understand the time taken for the proposed methods for the training of models. We also want to explore how matrix factorization or SVD can be integrated with GRL to improve empirical performance and is it possible to extract the goodness from both methods and merge them. We also plan to work on data profiling for the proposed methods and existing literature to understand why one approach performs well on some datasets but does not perform equally well on other datasets. On the empirical investigation front, we also intend to benchmark these approaches on the Open Graph Benchmark \cite{hu2020open}.  We also plan to test out these approaches in tasks beyond recommendation systems such as graph-based formulations in NLP, social network modelling and graph applications in biology.

Through our work, we want to highlight that matrix factorization based methods still contribute as important baselines and should not be ignored in empirical benchmarking while making more advances in GRL or recommendation systems.

\bibliographystyle{IEEEbib}
\bibliography{main}

\begin{thebibliography}{10}

\bibitem{grover2016node2vec}
Aditya Grover and Jure Leskovec,
\newblock ``node2vec: Scalable feature learning for networks,''
\newblock in {\em Proceedings of the 22nd ACM SIGKDD international conference
  on Knowledge discovery and data mining}, 2016, pp. 855--864.

\bibitem{kipf2016semi}
Thomas~N Kipf and Max Welling,
\newblock ``Semi-supervised classification with graph convolutional networks,''
\newblock {\em arXiv preprint arXiv:1609.02907}, 2016.

\bibitem{van2017graph}
Rianne van~den Berg, Thomas~N Kipf, and Max Welling,
\newblock ``Graph convolutional matrix completion,''
\newblock {\em arXiv preprint arXiv:1706.02263}, 2017.

\bibitem{wang2019heterogeneous}
Xiao Wang, Houye Ji, Chuan Shi, Bai Wang, Yanfang Ye, Peng Cui, and Philip~S
  Yu,
\newblock ``Heterogeneous graph attention network,''
\newblock in {\em The World Wide Web Conference}, 2019, pp. 2022--2032.

\bibitem{hamilton2017inductive}
William~L Hamilton, Rex Ying, and Jure Leskovec,
\newblock ``Inductive representation learning on large graphs,''
\newblock in {\em Proceedings of the 31st International Conference on Neural
  Information Processing Systems}, 2017, pp. 1025--1035.

\bibitem{gonzalez2021predicting}
Guadalupe Gonzalez, Shunwang Gong, Ivan Laponogov, Michael Bronstein, and
  Kirill Veselkov,
\newblock ``Predicting anticancer hyperfoods with graph convolutional
  networks,''
\newblock {\em Human Genomics}, vol. 15, no. 1, pp. 1--12, 2021.

\bibitem{laponogov2021network}
Ivan Laponogov, Guadalupe Gonzalez, Madelen Shepherd, Ahad Qureshi, Dennis
  Veselkov, Georgia Charkoftaki, Vasilis Vasiliou, Jozef Youssef, Reza
  Mirnezami, Michael Bronstein, et~al.,
\newblock ``Network machine learning maps phytochemically rich “hyperfoods”
  to fight covid-19,''
\newblock {\em Human Genomics}, vol. 15, no. 1, pp. 1--11, 2021.

\bibitem{wu2021bridgedpi}
Yifan Wu, Min Gao, Min Zeng, Feiyang Chen, Min Li, and Jie Zhang,
\newblock ``Bridgedpi: A novel graph neural network for predicting drug-protein
  interactions,''
\newblock {\em arXiv preprint arXiv:2101.12547}, 2021.

\bibitem{liu2018heterogeneous}
Ziqi Liu, Chaochao Chen, Xinxing Yang, Jun Zhou, Xiaolong Li, and Le~Song,
\newblock ``Heterogeneous graph neural networks for malicious account
  detection,''
\newblock in {\em Proceedings of the 27th ACM International Conference on
  Information and Knowledge Management}, 2018, pp. 2077--2085.

\bibitem{qu2020category}
Xiaoru Qu, Zhao Li, Jialin Wang, Zhipeng Zhang, Pengcheng Zou, Junxiao Jiang,
  Jiaming Huang, Rong Xiao, Ji~Zhang, and Jun Gao,
\newblock ``Category-aware graph neural networks for improving e-commerce
  review helpfulness prediction,''
\newblock in {\em Proceedings of the 29th ACM International Conference on
  Information \& Knowledge Management}, 2020, pp. 2693--2700.

\bibitem{ragesh2021hetegcn}
Rahul Ragesh, Sundararajan Sellamanickam, Arun Iyer, Ramakrishna Bairi, and
  Vijay Lingam,
\newblock ``Hetegcn: heterogeneous graph convolutional networks for text
  classification,''
\newblock in {\em Proceedings of the 14th ACM International Conference on Web
  Search and Data Mining}, 2021, pp. 860--868.

\bibitem{yao2019graph}
Liang Yao, Chengsheng Mao, and Yuan Luo,
\newblock ``Graph convolutional networks for text classification,''
\newblock in {\em Proceedings of the AAAI conference on artificial
  intelligence}, 2019, vol.~33, pp. 7370--7377.

\bibitem{zhang2020every}
Yufeng Zhang, Xueli Yu, Zeyu Cui, Shu Wu, Zhongzhen Wen, and Liang Wang,
\newblock ``Every document owns its structure: Inductive text classification
  via graph neural networks,''
\newblock {\em arXiv preprint arXiv:2004.13826}, 2020.

\bibitem{linmei2019heterogeneous}
Hu~Linmei, Tianchi Yang, Chuan Shi, Houye Ji, and Xiaoli Li,
\newblock ``Heterogeneous graph attention networks for semi-supervised short
  text classification,''
\newblock in {\em Proceedings of the 2019 Conference on Empirical Methods in
  Natural Language Processing and the 9th International Joint Conference on
  Natural Language Processing (EMNLP-IJCNLP)}, 2019, pp. 4823--4832.

\bibitem{wang2019neural}
Xiang Wang, Xiangnan He, Meng Wang, Fuli Feng, and Tat-Seng Chua,
\newblock ``Neural graph collaborative filtering,''
\newblock in {\em SIGIR}, 2019.

\bibitem{zhao2017meta}
Huan Zhao, Quanming Yao, Jianda Li, Yangqiu Song, and Dik~Lun Lee,
\newblock ``Meta-graph based recommendation fusion over heterogeneous
  information networks,''
\newblock in {\em Proceedings of the 23rd ACM SIGKDD International Conference
  on Knowledge Discovery and Data Mining}, 2017, pp. 635--644.

\bibitem{he2020lightgcn}
Xiangnan He, Kuan Deng, Xiang Wang, Yan Li, Yongdong Zhang, and Meng Wang,
\newblock ``Lightgcn: Simplifying and powering graph convolution network for
  recommendation,''
\newblock in {\em Proceedings of the 43rd International ACM SIGIR conference on
  research and development in Information Retrieval}, 2020, pp. 639--648.

\bibitem{sunmulti}
Jianing Sun and Yingxue Zhang,
\newblock ``Multi-graph convolutional neural networks for representation
  learning in recommendation,''
\newblock .

\bibitem{zhao2019t}
Ling Zhao, Yujiao Song, Chao Zhang, Yu~Liu, Pu~Wang, Tao Lin, Min Deng, and
  Haifeng Li,
\newblock ``T-gcn: A temporal graph convolutional network for traffic
  prediction,''
\newblock {\em IEEE Transactions on Intelligent Transportation Systems}, vol.
  21, no. 9, pp. 3848--3858, 2019.

\bibitem{wu2019simplifying}
Felix Wu, Amauri Souza, Tianyi Zhang, Christopher Fifty, Tao Yu, and Kilian
  Weinberger,
\newblock ``Simplifying graph convolutional networks,''
\newblock in {\em International conference on machine learning}. PMLR, 2019,
  pp. 6861--6871.

\bibitem{ungar1998clustering}
Lyle~H Ungar and Dean~P Foster,
\newblock ``Clustering methods for collaborative filtering,''
\newblock in {\em AAAI workshop on recommendation systems}. Menlo Park, CA,
  1998, vol.~1, pp. 114--129.

\bibitem{li2019improved}
Xiaofeng Li and Dong Li,
\newblock ``An improved collaborative filtering recommendation algorithm and
  recommendation strategy,''
\newblock {\em Mobile Information Systems}, vol. 2019, 2019.

\bibitem{debnath2008feature}
Souvik Debnath, Niloy Ganguly, and Pabitra Mitra,
\newblock ``Feature weighting in content based recommendation system using
  social network analysis,''
\newblock in {\em Proceedings of the 17th international conference on World
  Wide Web}, 2008, pp. 1041--1042.

\bibitem{choi2012hybrid}
Keunho Choi, Donghee Yoo, Gunwoo Kim, and Yongmoo Suh,
\newblock ``A hybrid online-product recommendation system: Combining implicit
  rating-based collaborative filtering and sequential pattern analysis,''
\newblock {\em electronic commerce research and applications}, vol. 11, no. 4,
  pp. 309--317, 2012.

\bibitem{klema1980singular}
Virginia Klema and Alan Laub,
\newblock ``The singular value decomposition: Its computation and some
  applications,''
\newblock {\em IEEE Transactions on automatic control}, vol. 25, no. 2, pp.
  164--176, 1980.

\bibitem{hansen1990truncated}
Per~Christian Hansen,
\newblock ``Truncated singular value decomposition solutions to discrete
  ill-posed problems with ill-determined numerical rank,''
\newblock {\em SIAM Journal on Scientific and Statistical Computing}, vol. 11,
  no. 3, pp. 503--518, 1990.

\bibitem{rendle2012bpr}
Steffen Rendle, Christoph Freudenthaler, Zeno Gantner, and Lars Schmidt-Thieme,
\newblock ``Bpr: Bayesian personalized ranking from implicit feedback,''
\newblock {\em arXiv preprint arXiv:1205.2618}, 2012.

\bibitem{kingma2014adam}
Diederik~P Kingma and Jimmy Ba,
\newblock ``Adam: A method for stochastic optimization,''
\newblock {\em arXiv preprint arXiv:1412.6980}, 2014.

\bibitem{liang2018variational}
Dawen Liang, Rahul~G Krishnan, Matthew~D Hoffman, and Tony Jebara,
\newblock ``Variational autoencoders for collaborative filtering,''
\newblock in {\em Proceedings of the 2018 world wide web conference}, 2018, pp.
  689--698.

\bibitem{rao2015collaborative}
Nikhil Rao, Hsiang-Fu Yu, Pradeep~K Ravikumar, and Inderjit~S Dhillon,
\newblock ``Collaborative filtering with graph information: Consistency and
  scalable methods,''
\newblock in {\em Advances in Neural Information Processing Systems}, 2015, pp.
  2107--2115.

\bibitem{he2017neural}
Xiangnan He, Lizi Liao, Hanwang Zhang, Liqiang Nie, Xia Hu, and Tat-Seng Chua,
\newblock ``Neural collaborative filtering,''
\newblock in {\em Proceedings of the 26th international conference on world
  wide web}, 2017, pp. 173--182.

\bibitem{alexopoulos2020two}
Athanasios Alexopoulos, Georgios Drakopoulos, Andreas Kanavos, Phivos Mylonas,
  and Gerasimos Vonitsanos,
\newblock ``Two-step classification with svd preprocessing of distributed
  massive datasets in apache spark,''
\newblock {\em Algorithms}, vol. 13, no. 3, pp. 71, 2020.

\bibitem{sun2015spark}
Zhongyi Sun, Fengke Chen, Mingmin Chi, and Yangyong Zhu,
\newblock ``A spark-based big data platform for massive remote sensing data
  processing,''
\newblock in {\em International Conference on Data Science}. Springer, 2015,
  pp. 120--126.

\bibitem{hu2020open}
Weihua Hu, Matthias Fey, Marinka Zitnik, Yuxiao Dong, Hongyu Ren, Bowen Liu,
  Michele Catasta, and Jure Leskovec,
\newblock ``Open graph benchmark: Datasets for machine learning on graphs,''
\newblock {\em arXiv preprint arXiv:2005.00687}, 2020.

\end{thebibliography}

\end{document}